\documentstyle[aps]{revtex}
  \begin{document}
  
  \title {Black Hole Entropy from Loop Quantum Gravity}
  \author {Carlo Rovelli \thanks{e-mail: rovelli@pitt.edu} \\
  {\it Department of Physics and Astronomy, University of 
  Pittsburgh, 
  Pittsburgh, Pa 15260, USA}}
  \date{\today}
  \maketitle
  
  \begin{abstract}

\noindent We argue that the statistical entropy relevant for the 
thermal interactions of a black hole with its surroundings is (the 
logarithm of) the number of quantum microstates of the hole which 
are  distinguishable from the hole's exterior, and which correspond 
to a given hole's macroscopic configuration.  
We compute this number explicitly from first principles, for a 
Schwarzschild black hole, using nonperturbative quantum gravity in 
the loop representation. We obtain a black hole entropy proportional 
to the area, as in the Bekenstein-Hawking formula. 
  \end{abstract}

\vskip1cm
  
\noindent  In this letter, we present a derivation of the 
  Bekenstein-Hawking expression for the entropy of a  
  Schwarzschild black hole of surface area $A$ 
  \cite{wald}
  \begin{equation}
                 S = c\ {k\over \hbar G} \ A               
  \label{uno}
  \end{equation}
  (where $c$ is a constant of the order of unity, $G$ the Newton 
  constant, $k$ the Boltzman constant, and we have put the speed of 
  light equal to 1) via a statistical mechanical computation from a 
  full theory of quantum gravity \cite{tutti}.   We use the 
  loop representation of quantum gravity \cite{loops}, and, in 
  particular, we make use of the spectrum of the area operator 
  recently computed in loop quantum gravity \cite{area}.    
  This work is strongly influenced by (but conceptually different  
  from) a number of ideas presented by Krasnov in \cite{kirill}.
  
  Consider a system containing, among other components, a 
  non-rotating and non-charged black hole of mass $M$ (and therefore   
  surface area $A=16\pi G^2M^2$).  We are interested in the 
  (statistical) thermodynamics of such system.  The
  macroscopic state of the black hole is determined 
  by the single parameter $M$. However, there may be a large 
  number of microstates corresponding to the same macroscopic 
  configuration.  The number of such microstates  
  determines the entropy to be associated to the black hole in 
  analyzing its thermal interactions with the surroundings -- 
  in the same way in which the number of microstates of any given 
  subsystem determines the entropy of the subsystem at any given 
  macroscopic equilibrium state \cite{york}.  
  Taking quantum theory into account, such an entropy will
  be determined by the number of (linearly independent) quantum    
  states $|\psi_i\rangle$ that correspond to  
  the given macroscopic configuration of the hole.  

  The precise specification of these states is crucial.  
  First, macroscopical spherical symmetry 
  does not imply that individual statistical, or quantum, 
  fluctuations be spherically symmetric as well; therefore the 
  quantum microscopic configurations we have to consider  
  do not need to be individually spherically symmetric.   
  Second, only the configurations of the hole 
  itself, and not the configurations of the surrounding 
  geometry, affect the hole entropy. 
   Indeed, the surrounding geometry will have 
  many microscopic configurations corresponding to the 
  same ``macroscopic metric'', but this multiplicity will 
  contribute to an eventual entropy of the gravitational 
  field --of the gravitational radiation-- not to 
  the entropy to be associated to the hole.  Thus, we must 
  focus on quantum states of the hole.   Finally, what we are 
  considering is the thermodynamical behavior of a system 
  containing the hole. This behavior cannot be affected 
  by the hole's interior.  Indeed, the black hole interior 
  may be in an infinite number of states. For instance, the 
  black hole interior may (in principle) be given, say, by a Kruskal 
  extension, so that on the other side of the hole there is another 
  huge universe (maybe spatially compact, if not for the hole) 
  possibly with millions of galaxies.    The number of those internal 
  states cannot affect the interaction of the hole with its 
  surroundings. In other words, we are only interested in 
  configurations of the hole {\it that are distinguishable from the 
  exterior of the hole}.  From the exterior, the  
  hole is completely determined by the properties of its 
  surface.   Thus, the entropy relevant for the thermodynamical
  description of the thermal interaction of the hole with its 
   surroundings is determined by the states of the 
  quantum gravitational field (of the quantum geometry) on the 
  black hole surface.  
  
  An important point to take into account (missed in 
  an earlier stage of this work \cite{seminar}) is that for an
  external observer different  
  regions on the black hole surface are distinguishable from 
  each other. This is indeed easy to 
  see: consider initial data for the Einstein
  equations given in an asymptotically flat space containing
  a horizon. Consider data corresponding to a non-spherical
  deformation of a spherically symmetric event horizon.  Imagine 
  that this deformation is located in a certain region of the horizon. 
  Then the future evolution of the field --for instance the radiation 
  that reaches future infinity-- depends on the location of the 
  deformation on the event horizon. 
  Thus we are interested in the quantum states of the geometry 
  on a surface $\Sigma$ of area $A$, where different regions of 
  $\Sigma$ are distinguishable from each other.  At this point 
  the problem is well defined, and can be translated into a 
  direct computation, provided that a quantum theory of geometry is 
  given \cite{isham}. 

 \vskip.3cm  

  In loop quantum gravity, the quantum states of the 
  gravitational field are represented by s-knots 
  \cite{spinnet}. An s-knot is an equivalence class under 
  diffeomorphisms of graphs immersed in space, carrying colors 
  on their links (corresponding to irreducible 
  representations of $SU(2)$), and colors on their vertices 
  (corresponding  to invariant couplings between such 
  representations).   The relation between s-knots and classical  
  geometries was explored  in \cite{weave}.
  If a surface $\Sigma$ is given, the geometry of $\Sigma$  
  is determined by the intersections of the s-knot with the 
  surface.  Given a quantum state and a surface, let $i=1...n\ $ 
  label such intersections, and $p_i$ be the color of the link 
  through $i$. Generically, no node of the graph will be on 
  the surface; here, we disregard the 
  ``degenerate'' cases in which a node falls on the surface.  
    Thus, the quantum geometry of the surface is characterized by  
  an n-tuple of $n$ colors $\vec p=(p_1, 
  ..., p_n)$, where $n$ is arbitrary.  In particular, it was 
  shown in \cite{area} that the total area of the surface 
  $\Sigma$ is
  \begin{equation}
            A=\sum_{i=1,n} 8\pi\hbar G\, \sqrt{p_i(p_i+2)}.
  \label{area}
  \end{equation}
   Recall that we are assuming that the points of $\Sigma$ are 
  distinguishable. Thus a quantity observable from 
   outside the hole is, say,  the 
  area associated to a region containing only the 
  intersections $i=1,..., l<n$. Its value is
  \begin{equation}
            A = \sum_{i=1,l}  8\pi\hbar G\, \sqrt{p_i(p_i+2)}.
  \end{equation}
  Therefore, the quantum geometry on the surface is determined 
  by the {\it ordered\ } n-tuples of integers $\vec p=(p_1, ..., p_n)$. 
  States labeled by different orderings of the same unordered n-tuple 
  are distinguishable for an external observer.   
  
  We are thus interested in the number of ordered n-tuples of 
  integers $\vec p$ such that the macroscopic geometry of the 
  surface is the geometry of the surface of the black hole. The 
  geometry of the surface of the Schwarzschild black hole is    
  characterized  by the total area $A$, and by the uniformity of the 
  distribution of the area over the surface. Uniformity is 
  irrelevant in a count of microscopic configurations, 
  because the number of configurations and the number of 
  uniform configurations are virtually the same for large 
  area (for the vast majority of random 
  configurations of air molecules in a room, the macroscopic 
  air density is uniform).  Thus, our task of counting 
  microscopic configurations is reduced to the task of 
  counting  the ordered n-tuples of integers $\vec p$ such that 
  (\ref{area}) holds. More precisely, we are interested in the 
  number of microstates (n-tuples $\vec p$) such that the l.h.s of    
(\ref{area}) is between $A$ and $A+dA$, where $A>>\hbar G$ and $dA$ 
 is much smaller than $A$, but still macroscopic. 
  
  Let 
  \begin{equation}
                 M = {A \over 8\pi \hbar G },
  \label{m}
  \end{equation}
  and let $N(M)$ be the number of ordered n-tuples $\vec p$, with 
  arbitrary $n$, such that
  \begin{equation}
            \sum_{i=1,n} \sqrt{p_i(p_i+2)} = M. 
  \label{part}
  \end{equation} 
  First, we over-estimate 
   $M(N)$ by approximating the l.h.s. of (\ref{part}) 
  dropping the $+2$ term under the square root. Thus, we want to   
 compute the number $N_+(M)$ of ordered n-tuples such that 
\begin{equation}
            \sum_{i=1,n} p_i = M. 
\label{piu}
  \end{equation}
  The problem is a simple exercise in combinatorics. It can 
  be solved, for instance, by noticing that if $(p_1, ..., p_n)$ 
is a partition of $M$ (that is, it solves (\ref{piu}) ), then  
$(p_1, ..., p_n,1)$ and $(p_1, ..., p_n+1)$ are partitions of 
$M+1$.  Since all partitions  of $M+1$ can be obtained in this
 manner, we have
\begin{equation}
            N_+(M+1)= 2 N_+(M). 
  \end{equation}
Therefore 
\begin{equation}
            N_+(M)= C\ 2^M. 
  \end{equation}
Where $C$ is a constant. In the limit of large $M$ we have
\begin{equation}
	\ln N_+(M) = (\ln 2)\ M. 
\end{equation}
Next, we under-estimate $M(N)$ by approximating (\ref{part}) as
\begin{equation}
	\sqrt{p_i(p_i+2)}=\sqrt{(p_i+1)^2-1} \approx (p_i+1).
\end{equation}
Thus, we wish to compute the number $N_-(M)$ of ordered n-tuples such that 
\begin{equation}
            \sum_{i=1,n} (p_i+1) = M. 
  \end{equation}
Namely, we have to count the partitions of $M$ in parts with 2 or more 
elements.  This problem can be solved by noticing that if $(p_1, ..., p_n)$ 
is one such partition of $M$ and $(q_1, ..., q_m)$ is one such partition 
of $M-1$, then $(p_1, ..., p_n+1)$ 
and $(q_1, ..., q_m,2)$ are partitions of $M+1$. All partitions of $M+1$ 
in parts with 2 or more elements can be obtained in this manner, therefore
\begin{equation}
            N_-(M+1)= N_-(M) +  N_-(M-1)  .
\label{mmm}
  \end{equation}
It follows that  
\begin{equation}
            N_-(M)= D a_+^M + E a_-^M
\label{apm} 
  \end{equation}
where $D$ and $E$ are constants and $a_\pm$ (obtained by inserting 
(\ref{apm}) in (\ref{mmm})) are  the two roots of the equation
\begin{equation}
	a_\pm^2=a_\pm+1.
\end{equation}
In the limit of large $M$ the term with the highest root dominates, and 
we have
\begin{equation}
	\ln N_-(M) = (\ln a_+)\ M = \ln{1+\sqrt{5}\over 2}\ M. 
\end{equation}
By combining the information from the two estimates, we conclude that 
\begin{equation}
	\ln N(M) = d\ M. 
\end{equation}
where
\begin{equation}
	  \ln {1+\sqrt{5}\over 2}  < d  < \ln 2  
\end{equation}
or
\begin{equation}
	0.48 <  d  < 0.69. 
\end{equation}
Since the integers $M$ are equally spaced, our computation yields 
immediately the density of microstates. Using (\ref{m}), the 
number $N(A)$ of microstates with area 
$A$ grows for large $A$ as
  \begin{equation}
            \ln N(A) = d\  {{A\over8\pi \hbar G}}
\label{n}
  \end{equation}
  We have argued above that the statistical entropy that controls 
the thermal properties of the hole in its interactions with its 
surroundings is determined by the number of  microstates of the 
quantum geometry of the hole surface distinguishable from the
 exterior of the hole, namely
\begin{equation}
	S(A) = k \ln N(A)
\label{s}
\end{equation}
Equations (\ref{n}) and (\ref{s}) 
yield
  \begin{equation}
            S(A) =  c \ {k\over \hbar G} \  A . 
  \end{equation}
  which is the Bekenstein-Hawking formula.
  The constant of proportionality that we have 
  obtained is 
  \begin{equation}
            c = {d \over 8 \pi }, 
  \end{equation}
  which is roughly $4 \pi$ times smaller than Hawking's value    
  $c={1\over  4}$. 
  
  Notice that the dynamics (the Hamiltonian) does not enter      
  our derivation directly.  However, it does enter indirectly 
  by singling out the states with given area $A$ as the ones with the 
  same energy $M$. This is the usual role of the Hamiltonian in the 
  microcanonical framework.  In particular, in a gravitational theory
  different from GR the relation between the hole's energy $M$ and  
  its area $A$ may be altered --or even lost.  Thus, the 
  relation between number of states and area is purely
  kinematical, by the relation between this number and the entropy
  (which is the number of states with the same energy) is theory      
  dependent \cite{tate}. 

  We leave a number of issues open (which may affect the 
  proportionality factor).  We have disregarded the 
  degenerate states in which a node falls over the surface. 
  Also, we have worked in the simplified setting of a black hole 
  interacting with an external system with given geometry, instead      
  of working with a fully generally covariant statistical mechanics 
  \cite{stat}. 

  Finally, we comment on the relation of our result with 
  Ref.\ \cite{kirill}. We learned the idea of associating entropy to 
  classical configurations of the geometry --seen as 
  macroscopical states-- from Krasnov \cite{k}. An earlier 
  attempt to realize this idea along the lines described here 
  failed, yielding an entropy proportional to the square root 
  of the area \cite{seminar}.  A crucial breakthrough in 
  \cite{kirill} was the intuition that intersections are 
   distinguishable. In \cite{kirill}, however, the setting of the 
  problem is substantially different than ours: internal 
  configurations of the black hole are considered, the 
  Bekenstein entropy-bound conjecture and the holographic 
  conjecture are invoked in order to justify the counting 
  considered.  Here, we do not need those hypotheses. 
  Furthermore, the entropy-area 
  proportionality is derived in \cite{kirill} by means of an elegant
  but complicated argument involving a phase transition in a  
  fictitious auxiliary statistical system, while here the 
  combinatorial computation is performed explicitly.  
  
  In summary: we have argued that the black hole entropy 
  relevant for the hole's thermodynamical interaction with its 
  surroundings is the number of the quantum 
  microstates of the hole which are distinguishable from the 
  exterior of the hole; we have counted such microstates using loop 
  quantum gravity. We have obtained that the entropy is 
  proportional to the area, as in the Bekenstein-Hawking 
  formula, but with a different numerical proportionality 
  factor. 
    
  \vskip1cm
  
  I thank Abhay Ashtekar, Riccardo Capovilla, Luis Lehner,  Ted 
  Newman,  Jorge Pullin, Lee Smolin, and especially Kirill Krasnov  
  and Ranjeet Tate, for very stimulating discussions.

  \end{document}